\documentclass[twocolumn,prl,aps,preprintnumbers,amsmath,amssymb]{revtex4}
\usepackage[dvips]{graphicx}
\usepackage{dcolumn}
\usepackage{bm}
\usepackage{color}
\begin{document}

\title{Growth of thin graphene layers on stacked SiC surface in ultra high vacuum}

\author{C. \c{C}elebi, C. Yan{\i}k, A. G. Demirkol, and \.{I}smet \.{I}. Kaya}

\affiliation {Faculty of Engineering and Natural Sciences, Sabanci University, Tuzla, 34956 Istanbul, Turkey}

\date{\today}
\begin{abstract}
We demonstrate a technique to produce thin graphene layers on C-face of SiC under ultra high vacuum conditions. A stack of two SiC substrates comprising a half open cavity at the interface is used to partially confine the depleted Si atoms from the sample surface during the growth. We observe that this configuration significantly slows the graphene growth to easily controllable rates on C-face SiC in UHV environment. Results of low-energy electron diffractometry and Raman spectroscopy measurements on the samples grown with stacking configuration are compared to those of the samples grown by using bare UHV sublimation process.
\end{abstract}

\maketitle

Recent studies have stimulated a great interest in the controllable production of graphene due to its extraordinary two dimensional electronic properties \cite{Heer2007,Berger2004,Novoselov2004,Geim2007,Berger2006}. Epitaxial growth on SiC was proposed to be one of the most suitable methods for obtaining coherent and large-scale thin graphene templates which are compatible with the existing Si based electronic device fabrication technology \cite{Berger2004}. It is known that graphene forms in a self-assembled manner at temperatures above 1200 $^\circ$C both on Si-face (0001) and C-face (000$\bar{1}$) surfaces of SiC crystal by vacuum sublimation process. However, extremely high sublimation rate of Si atoms in a vacuum environment leads to the formation of crystalline defects and make it difficult to control the number of graphene layers especially on the C-face SiC surface \cite{Bommel1975,Mallet2007,Emtsev2008}.

It has been demonstrated that reducing the Si sublimation rate generates uniform epitaxial graphene with better thickness control. There are several approaches to suppress the depletion of Si atoms from SiC surface: annealing SiC in a graphite enclosure inside a high vacuum furnace \cite{Heer2011}, in a vapor phase silane environment \cite{Tromp2009} or in argon atmosphere \cite{Emtsev2009,Tedesco2010}. Carrier mobility of epitaxial graphene grown by using such sublimation control methods ranges from 10$^{3}$ cm$^{2}$V$^{-1}$s$^{-1}$ on the Si-face to 10$^{5}$ cm$^{2}$V$^{-1}$s$^{-1}$ on the C-face.

In this letter, we demonstrate an alternative technique that significantly reduces the graphene formation rate on the C-face surface of SiC still in ultra high vacuum (UHV) environment. Growth of thin layer epitaxial graphene is achieved by partially confining the Si vapor at the interface between two SiC substrates with a half open cavity in between. At high temperatures (around 1500 $^\circ$C), the cavity in between the two faces of the SiC stack suppresses the escape rate of the sublimated Si atoms from the sample surface and provides relatively high Si partial pressure. This local enhancement of Si vapor suppresses the rate of Si depletion from the SiC surface and hence leads to a reduced graphene growth rate compared to that grown by bare UHV sublimation process.

\begin{figure}[thb]
  \includegraphics[width=0.50\textwidth]{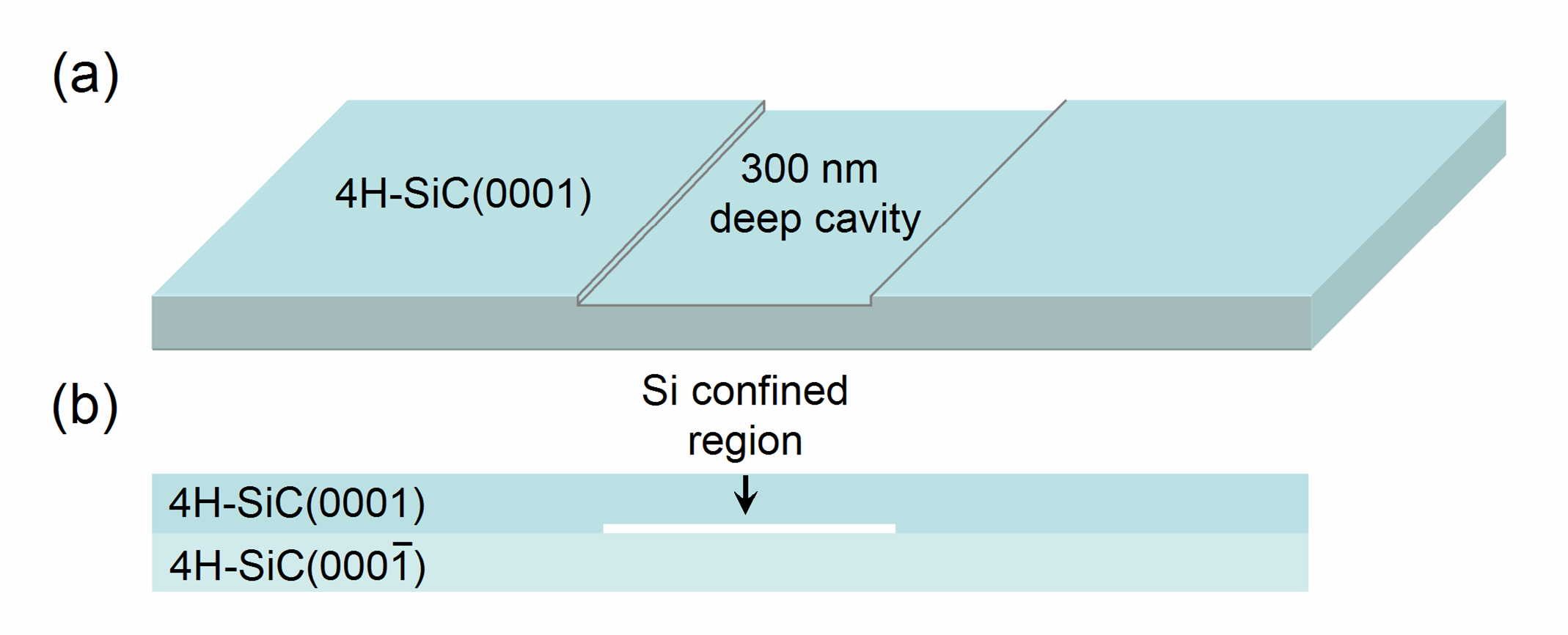}
\caption{(Color online). Schematic illustrations of (a) Si-face 4H-SiC capping substrate comprising a 300 nm deep cavity and (b) the side view of the stack: C-face SiC primary sample capped by Si-face 4H-SiC wafer.}
\label{fig1}
\end{figure}

For the experiments, we used 250 $\mu$m thick on-axis and n-type Si-face and C-face 4H-SiC wafers with atomically flat surfaces from NovaSiC. The wafers were diced into 3$\times$10 mm$^{2}$ rectangular samples and cleaned chemically. The native oxide layer on the sample is removed in diluted HF solution prior to loading into the UHV chamber which has a base pressure of P ${<}$ 1 $\times$ 10$^{-10}$ mbar. The samples were annealed in UHV by direct current heating during which the temperature is measured and controlled with 1 $^\circ$C resolution. As the capping substrate we used a SiC with the same dimensions, but with a 300 nm deep, 3 mm x 3 mm cavity on its Si-face as shown in Fig.~\ref{fig1}(a). Before capping on the primary sample, it was annealed separately in UHV for about 15 min. at 1430 $^\circ$C. When placed on the C-face surface of SiC wafer, the cavity on the cap provides a well defined 300 nm separation between its surface and the primary sample surface [Fig.~\ref{fig1}(b)]. The uniformity of the separation is verified by optical microscopy and optical interference patterns. This half open structure with very small aspect ratio (10$^{-4}$) keeps the Si atoms inside the cavity for elongated times in UHV environment.

\begin{figure}[thb]
  \includegraphics[width=0.47\textwidth]{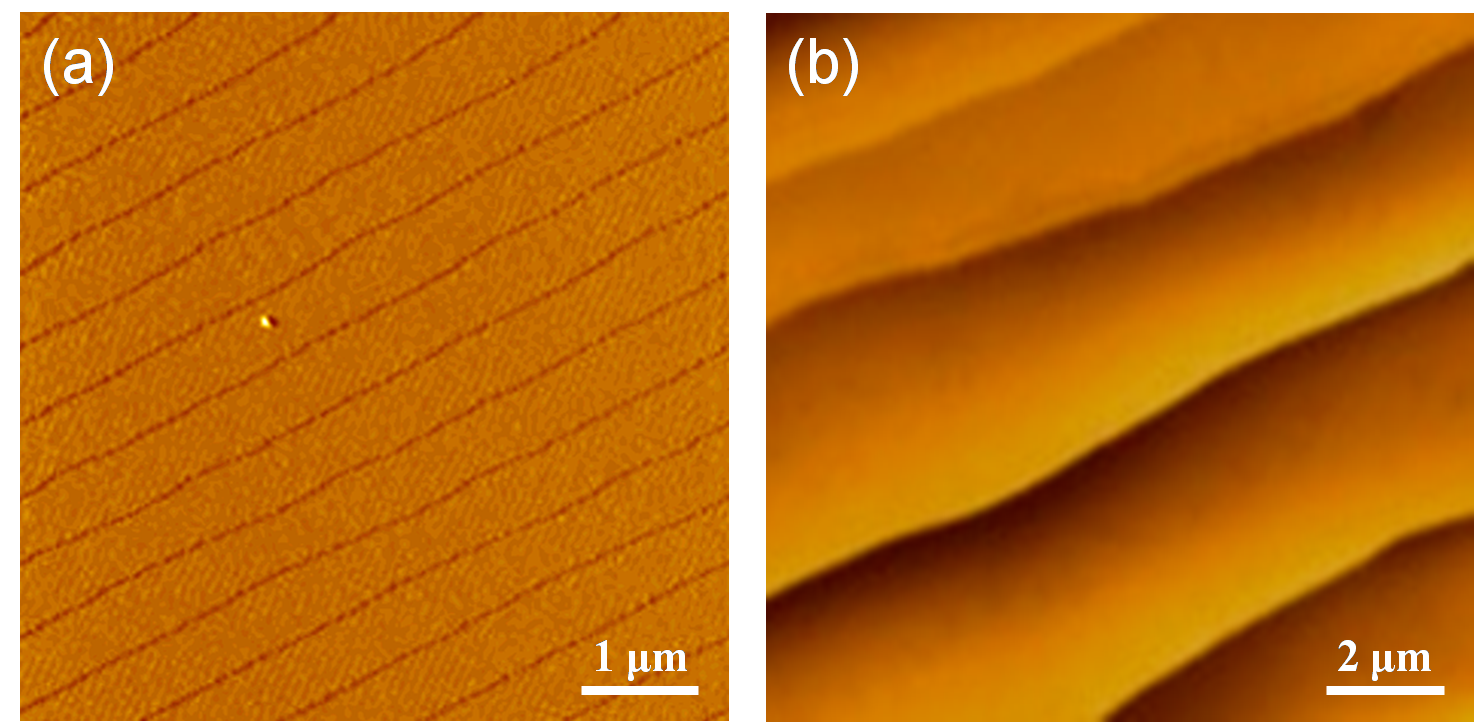}
\caption{(Color online). Tapping mode AFM topography images acquired on (a) as-received and (b) UHV annealed C-face 4H-SiC substrate (sample-A) surfaces. The increase of the terrace dimensions is clearly seen from (a) to (b).}
\label{fig2}
\end{figure}

The stack was annealed for 20 min. at 1500 $^\circ$C in UHV for graphene growth. After splitting the stack, C-face of SiC surface, on which graphene was grown, was characterized by atomic force microscopy (AFM), low-energy electron diffractometry (LEED) and Raman spectroscopy techniques. The LEED and Raman spectroscopy data are compared to those of another C-face SiC sample that was annealed at 1500 $^\circ$C for 5 min. in UHV with no capping substrate. For simplicity, samples annealed with and without the capping substrate were denoted as sample-A and sample-B, respectively. The alteration of the surface morphology of sample-A is analyzed by AFM imaging the as-received and the annealed surface of sample-A [Fig.~\ref{fig2}]. In the AFM scan of as-received sample, well ordered surface preparation induced 0.5 - 0.6 $\mu$m wide atomically flat terraces due to $\thicksim$0.1$^\circ$ miscut are seen. We observe that the terraces broaden to 4 - 6 $\mu$m after annealing, similar to those reported in Ref. \cite{Emtsev2009}.

\begin{figure}[thb]
  \includegraphics[width=0.47\textwidth]{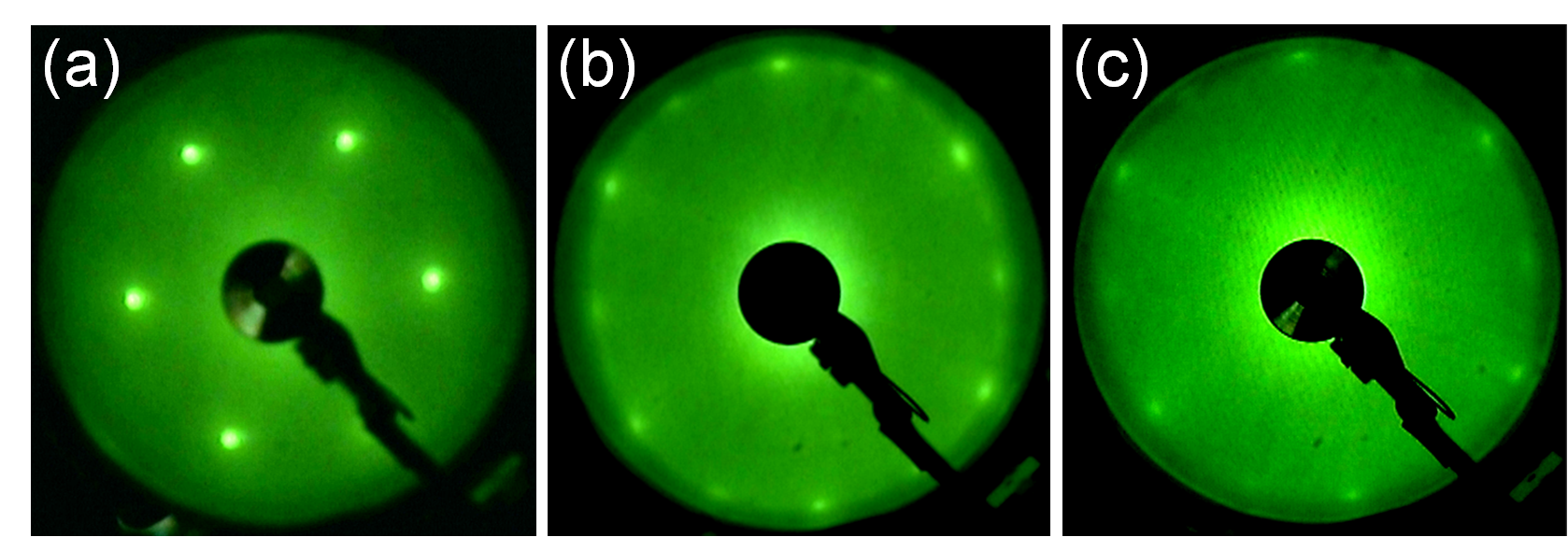}
\caption{(Color online). LEED patterns for as-received (a) and for graphitized C-face 4H-SiC samples which were annealed in UHV either (b) without (sample-B) or (c) with (sample-A) capping substrate. The LEED patterns correspond to the first order diffractions and were acquired at the incident electron energies of 100, 95 and 67 eV for (a), (b) and (c), respectively.}
\label{fig3}
\end{figure}

Epitaxial formation of graphene on the substrates annealed in UHV with both methods is determined by LEED measurements as shown in Figure 3. Hexagonally oriented structure with sharp spots seen in Fig.~\ref{fig3}(a) corresponds to the 6-fold (1$\times$1) diffraction pattern of a bare SiC surface. LEED pattern of the sample-B [Fig.~\ref{fig3}(b)] displays bright spots rotated 30$^\circ$ with respect to the bare SiC LEED pattern, and diffused arcs in between them. Here the bright spots correspond to the (1$\times$1) diffraction pattern of graphite and the arcs are due to the graphene layers rotationally disordered on the C-face surface of SiC. These observations are in good agreement with the previous reports \cite{Heer2011,Hass2006}. Sample-A also exhibits a low intensity 6-fold diffraction pattern [Fig.~\ref{fig3}(c)], as in the LEED pattern of sample-B. However, the diffused arcs in between the primary spots almost totally vanish for sample-A. These results are consistent with the existence of very small number of graphene layers on C-face SiC \cite{Heer2011}. The pattern is resolved only at the incident beam energy of 67 eV and entirely disappeared at higher orders, which also points out the formation of single or few layers.

\begin{figure}[thb]
  \includegraphics[width=0.47\textwidth]{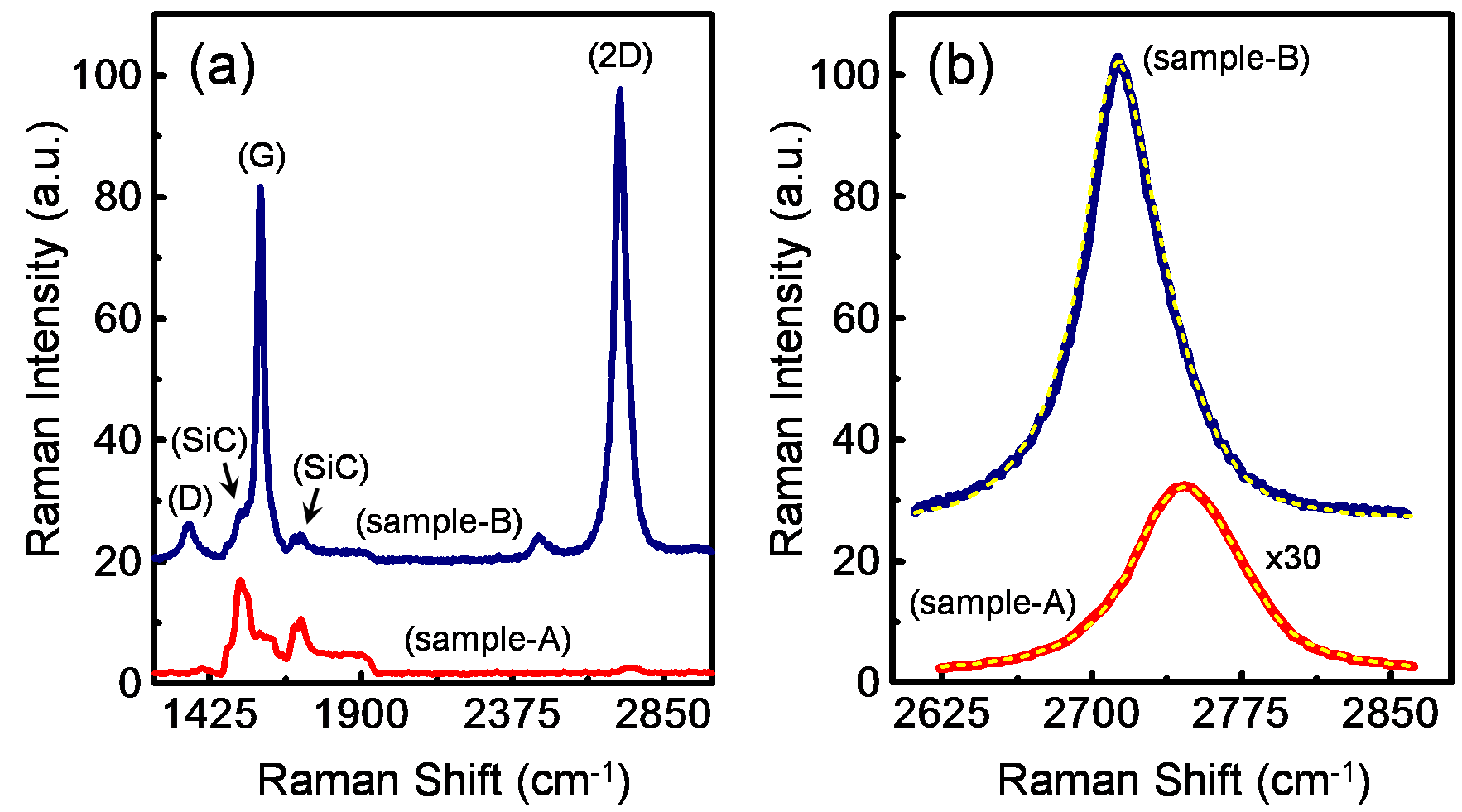}
\caption{(Color online). (a) Raman spectra acquired on the C-face 4H-SiC sample annealed in UHV with or without using the capping substrate. (b) Expanded views of the 2D band peaks in spectra (a) and the corresponding Lorentzian fits (dashed lines). The curves are offset along vertical axis for clarity.}
\label{fig4}
\end{figure}

Further analysis of the samples was done by Raman spectroscopy measurements performed in ambient conditions by using a green laser with excitation wavelength of 514 nm. The Raman spectra for both sample-A and sample-B are displayed in Fig.~\ref{fig4}(a). For sample-A, the measurements are performed on the stepped terraces seen in the AFM image [Fig.~\ref{fig2}(b)]. D, G and 2D bands of graphene are clearly resolved in the spectra together with the SiC induced background signals. Compared to sample-B, the trace for sample-A has strongly attenuated G and 2D peaks, nevertheless has relatively enhanced SiC induced features. These observations are consistent with very thin graphene layers grown on sample-A surface \cite{Ni2008}. The D band originates from the breakdown of the wavevector selection rule and reveals the presence of crystalline defects in the graphene matrix. The weak D band peak intensity seen in the Raman spectrum of sample-A implies low defect concentration in the graphene layers on this sample.

The clearly visible 2D bands of both spectra are smooth and have no extra features, which mean they can be best fitted to a single Lorentzian function as shown in Fig.~\ref{fig4}(b). The smooth shape of both curves implies that the stacking of graphene in these samples is not Bernal but, rather, turbostratic. This result is supported by our LEED measurements and consistent with the literature \cite{Hass2006}. Nearly unchanged position of the G band (1584 cm$^{-1}$ for sample-A and 1585 cm$^{-1}$ for sample-B) indicates that the graphene layers on both substrates are under the influence of similar amount of strain \cite{Ni2008}. It was shown that rotational stacking disorientation of epitaxial graphene on C-face SiC leads to a weak bonding of the first layer with the underlying substrate as well as to a strain relaxation along the graphene stacks \cite{Strudwick2011}. On the other hand, for sample-A, the measured D band ($\thicksim$1402 cm$^{-1}$) and 2D band ($\thicksim$2745 cm$^{-1}$) peak positions are found to be shifted ($\thicksim$40 cm$^{-1}$ for D band and $\thicksim$33 cm$^{-1}$ for 2D band) toward higher frequencies with respect to those of sample-B. Because the G band frequencies of both samples are similar, the observed shift in D and 2D band peak positions can not be interpreted as a strain phenomenon. According to double resonant Raman scattering model \cite{Thomsen2000}, the peak positions of D and 2D bands in the spectrum have strong dependence on the electronic structure of the sample as well as the laser excitation energy. Since all the Raman data in our measurements were acquired at a fixed wavelength (514 nm) the contribution from the laser excitation can thus be excluded. The observed shift may arise due to the charge-transfer doping \cite{Yang2010} from our n-type doped substrate into overlying thin graphene layers which maintains a strong mutual electronic coupling. As the coupling is stronger for small number of layers, the D and 2D band peak positions are expected to deviate more from bulk Raman frequencies similar to that seen in the Raman spectrum of sample-A [Fig.~\ref{fig4}].

In conclusion, we demonstrated that in UHV the high formation rate of graphene layers on C-face SiC can be significantly reduced when it is capped by another SiC substrate. Controlled confinement of the sublimated Si atoms is achieved by forming a half open cavity at the interface between a stack of two SiC crystals. LEED and Raman spectroscopy measurements reveal that single or few layer graphene is formed on the capped sample when annealed at 1500 $^\circ$C for 20 min., compared to formation of graphitic thick films only in 5 min. on bare samples at the same temperature. The results confirm that confined Si vapor in the vicinity of sample surface causes a strong reduction in the growth rate of graphene on the C-face SiC substrate in clean UHV environment.

The authors would like to thank Walt de Heer from Georgia Institute of Technology for valuable discussions, Mustafa \c{C}ulha from Yeditepe University for Raman spectroscopy measurements and Ahmet Oral from Sabanci University for AFM measurements. This work was supported by the Scientific and Technological Research Council of Turkey (TUBITAK) under Project Grant No. TBAG-107T855.


\end{document}